\documentclass[trackchanges]{aastex7}

\newcommand\msini{M$\sin{i}$}
\newcommand\mearth{M$_{\oplus}$}
\newcommand\mjup{M$_{\mathrm{J}}$}

\usepackage{amsmath}
\usepackage[T1]{fontenc}

\begin{document}

\title{Evidence for a Peak at $\sim$0.3 in the Eccentricity Distribution of Typical Super-Jovian Exoplanets}

\author[0000-0002-3199-2888]{Sarah Blunt}
\affiliation{Department of Astronomy and Astrophysics, University of California, Santa Cruz, CA, USA}
\affiliation{Center for Interdisciplinary Exploration and Research in Astrophysics (CIERA), Northwestern University, Evanston, IL 60208, USA}
\email[show]{sarah.blunt.3@gmail.com}  

\author[0000-0003-0774-6502]{Jason Wang}
\affiliation{Center for Interdisciplinary Exploration and Research in Astrophysics (CIERA), Northwestern University, Evanston, IL 60208, USA}
\affiliation{Department of Physics and Astronomy, Northwestern University, Evanston, IL, USA}
\email{}

\author[0000-0001-5061-0462]{Ruth Murray-Clay}
\affiliation{Department of Astronomy and Astrophysics, University of California, Santa Cruz, CA, USA}
\email{}

\author[0000-0003-1212-7538]{Bruce Macintosh}
\affiliation{Department of Astronomy and Astrophysics, University of California, Santa Cruz, CA, USA}
\email{}

\author[0000-0003-3856-3143]{Ryan A. Rubenzahl}
\affiliation{Center for Computational Astrophysics, Flatiron Institute, 162 Fifth Avenue, New York, NY 10010, USA}
\email{}

\author[0000-0003-3504-5316]{B.J. Fulton}
\affiliation{NASA Exoplanet Science Institute/Caltech-IPAC, California Institute of Technology, Pasadena, CA 91125, USA}
\email{}

\begin{abstract}

In this study, we compute completeness-corrected occurrence rates of giant exoplanets as a function of mass, semimajor axis, and eccentricity, using the approximately uniform California Legacy Survey sample of RV-discovered planets published in \cite{Rosenthal:2021a}. We recover the previously-detected rise in occurrence with semimajor axis for both lower- and higher-mass subsets of the population out to $\sim$5~au. When restricting to planets with semimajor axes between 0.1 and 4.5~au (roughly speaking, the ``peak'' of giant planet occurrence), we find evidence for distinct eccentricity distributions for each of two mass sub-populations. Most strikingly, we observe a peak in the eccentricity distribution of super-Jovian planets (3-20~M$_{\rm J}$) at 0.3, which is apparent using two different parameterizations of the eccentricity distribution model. A hierarchical histogram model reveals that $\sim$92\% of posterior samples indicate an elevated occurrence rate of super-Jupiters with modest eccentricities (0.2--0.4) compared to lower or higher eccentricities (i.e. evidence for a moderate eccentricity ``peak''), and 99\% of samples indicate super-Jupiters with modest eccentricities are more common than those with lower eccentricities (i.e. evidence that moderate eccentricities are more common than low eccentricities). We use a truncated Gaussian model fit to pinpoint the location of the super-Jupiter eccentricity peak with more precision, finding a maximum a posterior (MAP) peak location of $e=0.3$. This low but elevated characteristic eccentricity could be the result of dynamically hot histories, perhaps involving a giant impacts phase. All analysis code for this project is publicly available on Zenodo (\dataset[doi 18089157]{https://zenodo.org/records/18089157}) and GitHub: \href{https://github.com/sblunt/eccentricities}{github.com/sblunt/eccentricities}.
\end{abstract}


\section{Introduction} 

The orbits of planets, particularly orbital eccentricities, encode information about their formation and subsequent dynamical histories. Over the past 30 years of exoplanet science, this fact has enabled us to shed light on how planets form and evolve. The overall eccentricity distribution of exoplanets discovered with the radial velocity (RV) method is consistent with a beta distribution peaking at e=0 \citep{Kipping:2013a}. This is generally interpreted as evidence that exoplanets form via core accretion (which produces circular orbits), with the ``dynamical tail'' out to higher eccentricities resulting from a zoo of potential dynamical processes: planet-planet scattering (\citealt{Rasio:1996a}, \citealt{Chatterjee:2008a}, \citealt{Juric:2008a}, \citealt{Marleau:2019a}), secular interactions with outer planets or stellar companions (\citealt{Fabrycky:2007a}, \citealt{Naoz:2013a}), excitation due to stellar fly-bys \citep{DeRosa:2019a}, and planet-disk interactions \citep{Ragusa:2018a}. Overall, in the mid-2010s, all evidence pointed to most planets being on \textit{circular} orbits; while eccentric planets were known to exist, they were thought to be the exception to the rule.

Over the past few years, exoplanet science have begun to study orbital eccentricities in different regimes, pointing to a slightly more complicated story than vanilla core accretion and occasional dynamical excitation. At the low mass extreme, \cite{gilbert:2025a} recently showed that for small (0.5-15 R$_{\oplus}$) transiting planets, average eccentricity increases with mass, with 16 M$_{\oplus}$ objects achieving an average eccentricity of $\sim$0.2. At the high mass extreme, \citet{Bowler:2020a} identified a broad distribution of eccentricities for imaged brown-dwarf mass objects which resembles that of stellar binaries at similar separations (see, e.g. \citealt{Hwang:2022Ma}) more than that of planets discovered with RVs or imaging. This result was corroborated by \cite{Nagpal:2023a} and \cite{doo:2023a} (although the latter point out that the distinctness of the two distributions depends on the mass boundary). The elevated eccentricities of high-mass imaged brown dwarfs was interpreted by \citet{Bowler:2020a} as evidence for different formation mechanisms, with planets forming via core accretion and brown dwarfs via dynamically ``hotter'' process such as gravitational collapse. 

% These examples (and others) illustrate how orbital eccentricities of populations can be used to ``back out'' the physics that assembles planetary systems. 

In this study, we characterize the population-level eccentricity distribution of ``typical'' giant planets, defined as planets 175-6000 M$_{\oplus}$ located at the ``peak'' of giant planet occurrence ($\sim$0.1-4.6~au; \cite{Fulton:2021a}). The eccentricity distribution of this population has historically been difficult to understand; at such separations, transit probabilities are small, and RV discoveries require long baselines (generally at least one orbital period, although not always; see e.g., \citealt{Blunt:2019a}). In addition, direct imaging has historically been limited to wider-separation giants (although the advent of interferometric detection is changing this; see \citealt{Winterhalder:2024a} and \citealt{Hinkley:2023a}). Microlensing, which is undeniably useful for determining occurrence rate statistics (\citealt{Mroz:2023a}), is incapable of measuring exoplanet eccentricities.

As long-baseline RV surveys (\citealt{Rosenthal:2021a}, \citealt{Wittenmyer:2020a}) mature, uniform samples of RV-derived giant planet eccentricities at the peak of occurrence are available for the first time. In this Letter, inspired by trends first identified by \citet{Frelikh:2019a}, we take advantage of the recent public release of the California Legacy Survey (CLS; \citealt{Rosenthal:2021a}) to derive completeness-corrected occurrence rates as a function of eccentricity, semimajor axis, and mass.

\cite{Weldon:2025a} recently pointed out, using exoplanet eccentricity measurements downloaded from the NASA Exoplanet Archive, a peak in the eccentricity distribution of super-Jupiter planets similar to the result we ultimately derive here. Our objective in this paper is to evaluate the significance of this peak, taking into account survey sensitivity, Poisson statistics, and other potential statistical biases. We ultimately corroborate their observation of the peak, and ask that future works cite their paper alongside this work when referencing the super-Jupiter eccentricity peak. 

\section{Method}

\subsection{Method in Theory: a Hierarchical Histogram}

Our method expands on a  formalism originally developed by \cite{Hogg:2010a} and \citet{Foreman-Mackey:2014a}, and recently applied by \cite{Fulton:2021a} and \cite{gilbert:2025a} (among others). In Appendix \ref{sec:appendix1}, we present our method in detail. For casual readers, the method can be understood as making a 3-dimensional histogram of planets in (e, $a$, M). This histogram, however, is special because: 1) it corrects for non-uniform survey sensitivity (``completeness''), 2) it takes into account uncertainties in individual planets' posteriors on  (e, $a$, M) that make their ``bin identity'' uncertain, 3) it accounts for counting statistics uncertainties by treating the entire observed sample as a realization of a Poisson distribution, and 4) it marginalizes over the uncertainties in inclination at a population level, allowing us to calculate occurrence rates in (e, $a$, M), even though RV measurements can only constrain (e, $a$, \msini{}).

\subsection{Method in Practice}

We used the publicly available California Legacy Survey (CLS) sample \citep{Rosenthal:2021a} planets as an approximately uniform, uninformed long-baseline ($>10$ year) sample of FGK stars to search for companions. Section 2 of \citep{Rosenthal:2021a} describes the sample selection in detail, including observing strategy and stellar sample characteristics. Figure 2 of that paper shows the distribution of RV baselines in the sample. The CLS survey observed 719 stars over a median RV baseline of 21 yrs, each star for a median of 71 observations. Sample selection was performed so as not to bias the survey toward or away from detecting companions (e.g. stellar metallically was purposefully not factored into sample selection, and stars were not chosen or excluded from the sample because of known transiting planets). 

In order to characterize the completeness of the CLS survey, $Q(\boldsymbol{\omega})$, we used a series of injection-recovery tests presented in \cite{Rosenthal:2021a} (see their Section 5.2)\footnote{These are available at \href{https://github.com/leerosenthalj/CLSI/tree/master/completeness}{https://github.com/leerosenthalj/CLSI/tree/master/completeness}.}. For each star in the sample, 3000 sets of injections were performed, drawing all orbital parameters randomly. These injection-recovery tests used the same search methodology and recovery criteria as the CLS survey itself for determining whether or not a signal qualifies as a planet (see Sections 5.1 and 5.2 of \citealt{Rosenthal:2021a} for details). This is important; if we were to modify the process used to search the data for planets in any way, we would want to modify the injection-recovery process in the same way. 

The results of these injection-recovery tests appear to contradict prevailing wisdom that completeness in RV surveys decreases only slightly with eccentricity (e.g. \citealt{Shen:2008a}). This trend has been explained in the past by the fact that as eccentricity increases and orbits become more difficult to sample in time, their RV amplitudes also become larger. For example, \cite{Shen:2008a} found, using a suite of injection-recovery tests with simulated data, that planets with eccentricities of 0.05 had a $\sim$45\% chance of detection in their simulations, while planets with eccentricities of 0.95 had a $\sim$35\% change of detection. Debunking the exact reasons for this difference are beyond the scope of this paper, but there are several important differences between the methodology used here and that of \cite{Shen:2008a}: 1) \citealt{Shen:2008a} used a $\chi^2$ minimization algorithm that used the correct (simulated) parameters as initial guesses. This likely greatly aided the minimization algorithm in finding the global minima; 2) \citealt{Shen:2008a} did not use a significance criteria to define whether or not a planet was recovered, whereas \citealt{Rosenthal:2021a} did. When \cite{Shen:2008a} did define a significance criteria for the recovery, that the recovered RV semiamplitude be greater than 1.8 $\sigma$ greater than 0, the completeness function they recover is in fact much steeper with eccentricity, lending credibility to this idea.

Regardless of \textit{why} we find different completeness patterns than \citealt{Shen:2008a}, it is important that the injection-recovery tests use the same logic to decide on a detection as the actual detection algorithm used for the data. It is possible that with a different detection algorithm, more (or different) planets would be recovered in the same survey, and the completeness function would look different. In other words, the completeness function itself is not just a product of the data, but also of the detection algorithm.

Given the results of these completeness tests, we constructed a survey-averaged model of completeness, shown in Figure \ref{fig:completeness}, by adding up all the recoveries in a given bin and dividing by all the injections. In theory, these completeness values should have their own associated uncertainties, because \cite{Rosenthal:2021a} performed a finite number of injections. However, in practice, these uncertainties are small enough to be ignored. To illustrate this point, consider that the minimum number of injections performed when splitting the space into one bin in $a$, two bins in \msini{} (ignoring the largest mass bin, corresponding to brown dwarfs, which is discussed in more detail in Section \ref{sec:binaries}), and five bins in $e$ was 284, with 111 recoveries. Assuming Poisson statistics, the completeness in this bin is $39\pm6$\%, which is well within the uncertainties of our derived occurrence rates (Section \ref{sec:results}). In the bin with the most injections, the completeness is known to within 0.5\%. As Figure \ref{fig:completeness} shows, completeness varies significantly as a function of eccentricity, and cannot be ignored when computing RV occurrence rates.

With our completeness model $Q(\boldsymbol{\omega})$ in hand, we then drew posterior samples in $P$, $K$, $\sqrt{e}\cos{\omega}$,  $\sqrt{e}\sin{\omega}$, and M$_*$ from the MCMC chains summarized in \cite{Rosenthal:2021a}, and shared with us by B.J. Fulton, (private communication), and transformed these to samples in $e$, \msini{}, $i$, and $a$. We then followed the method outlined in Appendix \ref{sec:appendix2} to resample these posterior samples to be consistent with priors uniform in $e$, $\log{M\sin{i}}$, and $\log{a}$. Following the discussion in \cite{Hogg:2010a}, we drew 50 posterior samples for each planet in our sample to then use for our hierarchical modeling (see the end of their Section 3 for discussion on how many posterior samples are needed to recover reliable hierarchical posteriors). We used \texttt{emcee} \citep{emcee} to sample the hierarchical likelihood (Equation \ref{eq:money}), using 100 walkers and running each chain for 500 burn-in and 500 production steps. Convergence was assessed by eye; trend plots are available at \href{https://github.com/sblunt/eccentricities}{https://github.com/sblunt/eccentricities} for each run published in this paper.

We are primarily interested in the eccentricities of giant planets at the peak of occurrence. We wanted to exclude the hot Jupiters, which likely have a unique dynamical history from the longer period planets in the sample, motivating a lower limit of $\sim$0.1 au for our results. In addition, we wanted to exclude incomplete orbits and trends from our analysis. Each star in the CLS survey has an RV baseline of at least 8 years (and most are FGK stars), which motivated an upper limit of $\sim$4~au. We defined eccentricity bins uniformly between 0 and 1. Beyond these constraints, we initially opted to use the same bin definitions as \cite{Fulton:2021a}; i.e. bins in $a$ log-uniformly spaced between 0.11~au and 4.56~au, and mass bins with limits (30, 300, 6000, 30,000) $M_{\oplus}$. Without performing a marginalization over inclination (i.e. computing occurrence rates as a function of \msini{} rather than mass; Section \ref{sec:appendix1}), these bin definitions revealed a strong eccentricity peak. Performing the marginalization, however, flattened the peak, which led us to believe that the true mass minimum cutoff for the peak was higher than 300 \mearth{}. We subsequently changed the mass limits to (30, 1000, 6000, 30,000), and recovered the eccentricity peak. 

\subsection{Treating Close Stellar Binaries \& the Brown Desert}
\label{sec:binaries}

One important limitation of the CLS sample \citep{Rosenthal:2021a} is the unknown sample selection with respect to stellar binaries. It is clear from raw numbers of detected stellar binaries in the sample that the \textit{observed} number of stellar binaries is much lower than the expectation from independent stellar binary statistics (see, e.g. \citealt{Grether:2006a}, who estimate that 11\% of FGK stars have a close stellar companion ($>0.08$ M$_{\odot}$)with a $<$5 yr orbital period; compare this to the 8 detections, out of 719 monitored stars, of objects with \msini{} > 20 M$_{\rm J}$ and a$<$4.6au. We can assume, extrapolating from trends shown in Figure \ref{fig:completeness}, that sensitivity to such companions should have been very nearly perfect, which leads to the inference that early designers of the CLS survey must have stopped RV monitoring of targets that showed signs of large RV variations (i.e. suspected binaries were ``thrown out'' of the sample). 

Although problematic overall for occurrence rate calculations, and especially for formulations like ours that marginalize over inclination, we can show based on external knowledge that the peak we detect is unlikely to be driven by contamination from face-on stellar binaries. To illustrate this, let's assume the worst case scenario: that 11\% of sun-like stars have a stellar binary companion with true mass 0.08 M$_{\odot}$. The expected number of objects with \msini{} values in the super-Jupiter planet range we define in this paper (1000-6,000 M$_{\oplus}$) is therefore:

\begin{equation}
    0.11 \times 719 \times (\cos(\sin^{-1}\left(\frac{1000 M_{\oplus}}{0.08 M_{\odot}}\right) - \cos(\sin^{-1}\left(\frac{6000 M_{\oplus}}{0.08 M_{\odot}}\right)) = 1.97
\end{equation} (recall that this is the \textit{worst case scenario}). 

We also know from independent studies that brown dwarfs (6,000 M$_{\oplus}$-0.08 M$_{\odot}$) in the separation range we study in this paper are extremely uncommon (a feature commonly called the ``brown dwarf desert''); \citealt{Grether:2006a} estimate the occurrence rate of objects in this range to be $<$1\%, and more modern studies (e.g. \citealt{Unger:2023a}, \citealt{Barbato:2023a}) corroborate this statistic (but see also \citealt{Wallace:2026a}, which shows preliminary evidence for a higher occurrence rate based on Gaia DR3 astrometric excess). The number of CLS-detected objects with \msini{} values in the brown dwarf desert regime is 2, as compared to the expected $\sim$0.86 from the $\sim$0.12\% occurrence rate derived for brown dwarfs within 5 au\footnote{Their inner semimajor axis limit is 1au, while ours is 0.1 au, so strictly speaking, their occurrence rates are not comparable to ours. However, assuming that occurrence rate of brown dwarfs rises with log(semimajor axis) (which appears reasonable given that there are more RV-detected objects with minimum masses in the brown dwarf range located 1-5au than 0.1-1au on exoplanet.eu, and detection sensitivity is likely worse at larger separations), we would not expect their occurrence rate to change by more than 100\% when including further in brown dwarfs.} by \citealt{Barbato:2023a} (see their Section 7). 

We opted to explicitly model the occurrence rates of objects in this mass range in order to explore the impact of potential contamination from face-on brown dwarfs in the desert on our super-Jupiter occurrence rates. It is possible that the objects in the brown dwarf \msini{} range of CLS sample are face-on binaries, so the occurrence rates we derive for objects in the brown dwarf desert should not be trusted on their own (i.e. they may be overestimated); however, the fact that the occurrence rate of brown dwarfs in our sample is likely \textit{over}estimated provides strong evidence that the eccentricity peak we observe is not driven by face-on brown dwarfs.

\subsection{A Note about Wide Stellar Binaries}
\label{sec:wide-binaries}

As noted in the section above, there is limited written information about the CLS sample selection with respect to stellar binaries, so it is unclear whether stars with known (at the time) imaged companions would have been selected out of the sample. This is particularly important for understanding the potential role of secular perturbations (see Section \ref{sec:implications}) in shaping the eccentricity trends we observe. An excellent avenue for future work would be identifying and examining the characteristics of the observed wide binary systems around CLS host stars. Comparing the characteristics of those that were found \textit{not} to host planets with those of the field population would help address this question empirically. 

\section{Results \& Interpretation}
\label{sec:results}

\subsection{Observed Trends}

Our major results are shown in Figures \ref{fig:money} and \ref{fig:ecc}, which display the same posterior samples in different ways. Figure \ref{fig:money} shows, for both sub- and super-Jovian planets, the probability of a planet in each mass range falling in a given semimajor axis and eccentricity bin. We chose to present these results in terms of probability, rather than raw occurrence, to make it easier to compare within a given mass bin (since higher-mass planets overall have a lower occurrence rate than lower-mass planets). For each posterior sample, these probabilities were calculated by: 1) computing the total occurrence (number of planets per star) across all bins, 2) dividing the occurrence rate in each bin by the total occurrence, yielding a probability. This process was repeated for each posterior sample, giving distributions over the probabilities of ending up in each bin. The values shown in Figure \ref{fig:money} represent median and 1$\sigma$ credible intervals of these probabilities. Note that even though each individual \textit{posterior sample} produces probabilities that add up to 100\%, the median values may not sum to exactly 100\%. Figure \ref{fig:ecc} shows 1D eccentricity distributions in each bin of mass and semimajor axis. In this figure, we show occurrence rates, which we did not convert to relative probabilities. 

The major trend we observe is a peak in the eccentricity distribution of super-Jupiter (1000 M$_{\oplus}<$M$<$6000 M$_{\oplus}$) planets at moderate eccentricities (0.2-0.4). 99\% of posterior samples yielded higher occurrence rates for super-Jupiters in the moderate eccentricity bin (0.2-0.4) than the low eccentricity bin (0-0.2), showing strong evidence that occurrence rises with eccentricity for super-Jupiter planets, while 92\%  yielded higher occurrence rates for super-Jupiters in the moderate eccentricity bin (0.2-0.4) than in \textit{both} the low eccentricity and moderate eccentricity bins (0-0.2 and 0.4-0.6, respectively). 

To reveal a more precise location of the eccentricity peak within the 0.2-0.4 eccentricity bin, we also performed a fit with a truncated Gaussian, rather than a histogram. Specifically, we: 1) restricted the parameter space of interest to either the intermediate- or high-mass bins, 2) redefined $\Gamma_{\theta}$, used in our likelihood calculation in Equation \ref{eq:money}, as a Gaussian function of eccentricity. To simplify the math, we also did not properly marginalize over inclination in this exercise, instead converting \msini{} to mass by multiplying each \msini{} posterior by a distribution that is uniform in $\cos{i}$. We then fit for the mean, amplitude, and standard deviation of the Gaussian. Uniform hyperpriors were applied to $\mu$, $\sigma$, and $A$, restricted between [0,1], [0,1], and [0,100], respectively. Results are shown in Figure \ref{fig:gaussian} for the low- and high-mass planet populations. The main takeaways from this exercise are: 1) as expected, the intermediate mass population shows no preference for a peak at nonzero eccentricities, and 2) the high mass population prefers a peak location of e=0.3. However, the truncated Gaussian model may not be an accurate representation of the overall distribution shape, and we encourage caution in using this model beyond this peak-finding exercise.

Another trend worth emphasizing is that the occurrence rate of \textit{very eccentric} (e$>$0.8) exoplanets is likely nonzero for super-Jupiters within our semimajor axis bounds. This is perhaps obvious, given that the number of planets detected in this area of parameter space is nonzero. However, it is worth stating explicitly: the eccentricity distribution of super-Jovian (1000 M$_{\oplus}<$M$<$6000~M$_{\oplus}$) exoplanets around 1~au is \textit{nonzero} at \textit{both} high and low eccentricities, and is therefore likely inconsistent with Rayleigh and beta distribution parametric forms.

\subsection{Robustness}

In this section, we present a few consistency checks we performed on our results. This list is by no means exhaustive, and we encourage the community to continue thinking about other effects that might be affecting our interpretation of the data.

\subsubsection{Comparison to \cite{Fulton:2021a}}

First, we directly compared the outputs of our model with that of \cite{Fulton:2021a}, simply by fixing the number of eccentricity bins equal to 1. These results are shown in Figure \ref{fig:bj-comp}, and can be directly compared to Figure 5 in that paper (although our model extends over a narrower semimajor axis region). We performed a fit in boxes of M$\sin{i}$ (directly comparable to the method of \citealt{Fulton:2021a}), and a fit in boxes of mass, marginalizing over inclination (see Appendix \ref{sec:appendix1}). As Figure \ref{fig:bj-comp} shows, we recover the same overall trends as \cite{Fulton:2021a}, including higher overall occurrence for sub-Jovian than super-Jovian planets, a ``jump'' in occurrence rate for super-Jupiters at $\sim$1~au, and an approximately monotonic increase in occurrence rate for both populations between 0.1 and 5~au. Performing the marginalization over inclination does not change these trends significantly, beyond slightly smoothing the occurrence rate changes with semimajor axis and widening error bars. Small differences between the results shown in Figure 5 of \cite{Fulton:2021a} and Figure \ref{fig:bj-comp} of this paper may be due to the fact that \cite{Fulton:2021a} used a Gaussian Process prior on the histogram heights, following \cite{Foreman-Mackey:2014a}, which has the effect of ``smoothing'' the occurrence rate distributions, while we chose not to perform this regularization. However, tracking down the exact cause of these minor discrepancies is beyond the scope of this paper. Overall, the two independent implementations agree well, which gives us confidence that our model is working as designed. 

\subsubsection{Stellar Activity?}

Next, we briefly investigated whether stellar activity could conceivably be affecting our results. The CLS survey was already vetted for potential false positive signals related to stellar activity (see Section 5.4 of \citealt{Rosenthal:2021a}), perhaps the most important of which is long-term stellar activity cycles (analogous to the Sun's 11 year cycle). \cite{Rosenthal:2021a} perform a series of vetting procedures designed to weed out stellar activity signals, for example by removing periodic signals that show strong correlation with the activity-sensitive core strengths of stellar Ca HK lines (``S-indices''). S-index/RV correlations on $\sim$decade timescales have been studied before (e.g. \citealt{Costes:2021a}), and are generally seen as a good but incomplete activity signature, meaning that many but not all activity signals that manifest in stellar RVs will also manifest at the same periods in S-index timeseries. In this study, we are chiefly concerned with whether the major trend we recover, that super-Jovian planets most often have eccentricities between 0.2-0.4, and that low-mass giant planets do not, could be explained by long-term activity cycles. For this to be true, long-term activity signals would need to be predominantly non-sinusoidal (i.e. they would need to mimic moderately eccentric Keplerian signals), and have larger amplitudes than planetary signals (which would show up in the lower-mass bin, following the expected monotonically decreasing eccentricity distribution). We can mostly debunk this idea by thinking about the scales required; of the M$\sin{i}$> 1000 M$_{\oplus}$ planets detected in the CLS sample (at any semimajor axis), the minimum radial velocity semi-amplitude is $\sim$50 m/s, and the median is $\sim$300 m/s. This is about an order of magnitude larger than the characteristic variability amplitude due to spots for solar-mass and solar-age stars ($\sim$5-10 m/s; \citealt{Saar:1997a}, \citealt{Luhn:2020a}), and the long-term RV variability of the Sun ($\sim$5 m/s; \citealt{Lanza:2016a}, \citealt{Klein:2024a}). Explaining a large portion of this population as due to activity-cycles would likely require the radial velocity scale of long-term activity cycles to be significantly larger than the short-term rotationally-modulated variability (this is true for the Sun, but not by 10x; \citealt{Lanza:2016a}), and also for these huge signals not to manifest in the S-indices. While we do not believe this explains our results, we encourage further encourage careful vetting of long-baseline RV surveys for activity-induced false positives.

\subsubsection{Two Circular Planets or One Eccentric Planet?}
\label{sec:2for1}

\cite{Wittenmyer:2019a} explored the anxiety-inducing possibility that, under realistic sampling circumstances for $\sim$decade-timescale radial velocity surveys, a single eccentric planet signal and a two-planet circular model can be indistinguishable (hereafter called the ``2-for-1'' effect). They found that highly eccentric (e$>$0.5) planets are very difficult to mimic with two circular Keplerian signals, but that moderately eccentric signals \textit{can} be mimicked by two circular Keplerians. \cite{Wittenmyer:2019a} performed a suite of injection-recovery tests to quantify the impact of this problem, and established five criteria that must be passed for a given single eccentric planet fit to be deemed plausible (see their Section 3). For each individual object in the high-mass bin, we checked that these criteria were satisfied, finding that the objects in the sample by and large pass with flying colors. Of the five \cite{Wittenmyer:2019a} criteria for a given eccentric planet to be deemed plausible, only one super-Jovian planet within our semimajor axis limits fails a single test (but passes the other four); see Figure \ref{fig:2for1}. We therefore find it unlikely that the 2-for-1 effect explains the moderate-eccentricity planet pileup for super-Jupiters.

\begin{figure*}
\includegraphics[width=\textwidth]{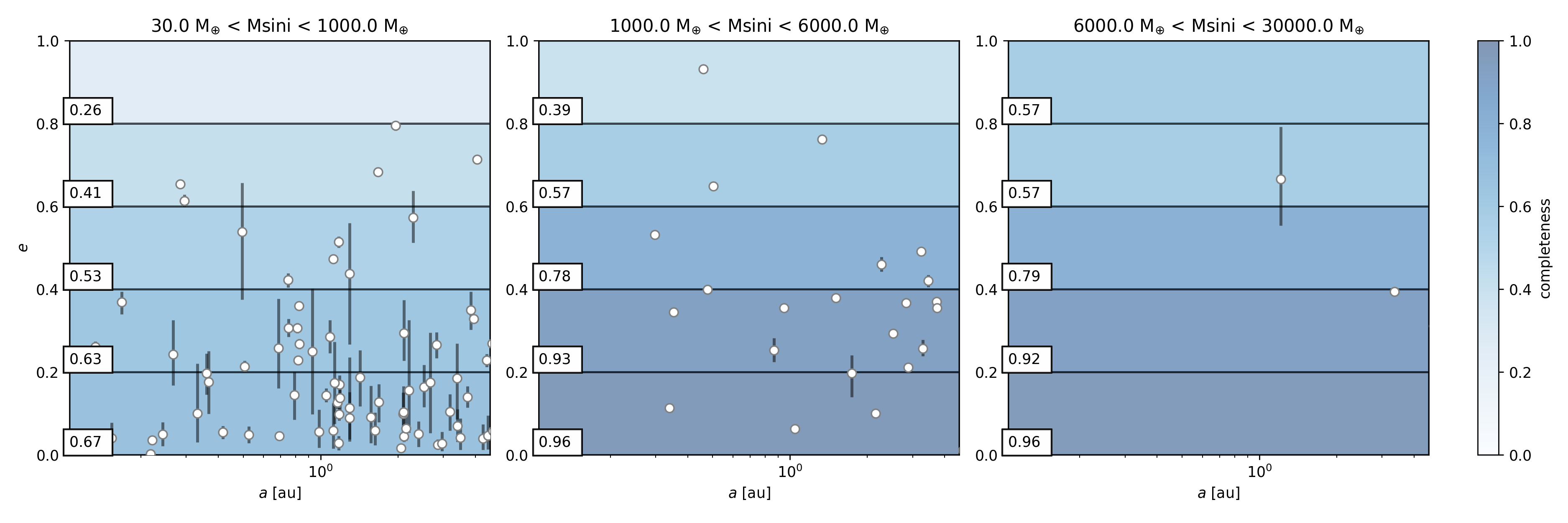}
\caption{\label{fig:completeness} Completeness model $Q(\boldsymbol{\omega})$ constructed for the CLS survey. Completeness is constructed using injection-recovery tests, and is shown in each survey box in blue. The CLS sample, shown as white points with error bars, is overplotted for reference. Individual planet are plotted either in the left or right panel based on their median \msini{}, but in the hierarchical model, a posterior sample from a given planet might fall in either the left or right panel (i.e. the mass space is continuous, not disjointed as it appears here). Takeaway: completeness worsens not only with decreasing \msini{}, but also with increasing eccentricity. This highlights the importance of including completeness in calculations of population-level eccentricity distributions using RVs.}
\end{figure*}

\begin{figure*}
\includegraphics[trim={3cm 0 2.5cm 0}, clip, width=\textwidth]{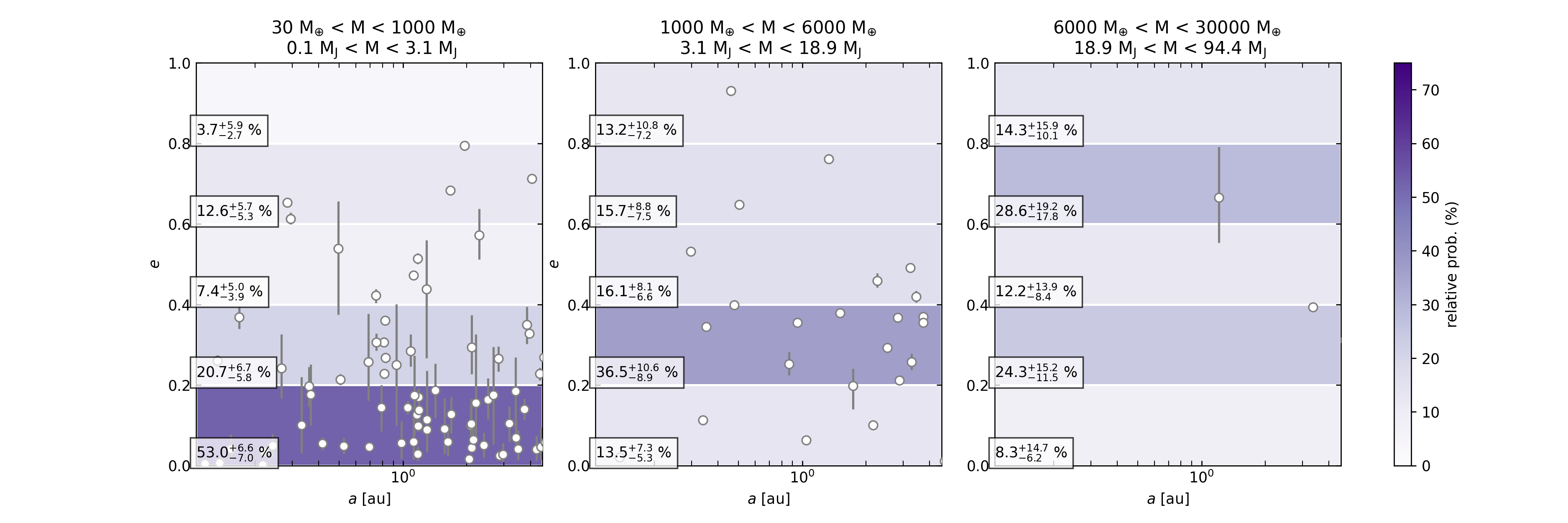}
\caption{Completeness-corrected occurrence rate estimates for sub-Jovian planets (left), super-Jovian planets (middle) and brown dwarfs (right), acknowledging the caveat that brown dwarf occurrence rates may be contaminated by face-on stellar binaries; see Section \ref{sec:binaries}. Color indicates overall occurrence rate in each ``box'' in the 3D parameter space, converted to a relative probability within a mass bin following the discussion in the text. Detected planets in the CLS sample are overplotted as white points with error bars. Takeaway: we recover the  expected monotonically decreasing eccentricity distribution of sub-Jovian planets. The occurrence of super-Jovian planets, however, shows a peak at moderate eccentricities. \label{fig:money}}
\end{figure*}

\begin{figure}
    \centering
    \includegraphics[width=0.5\linewidth, trim={0 12cm 0 0}, clip]{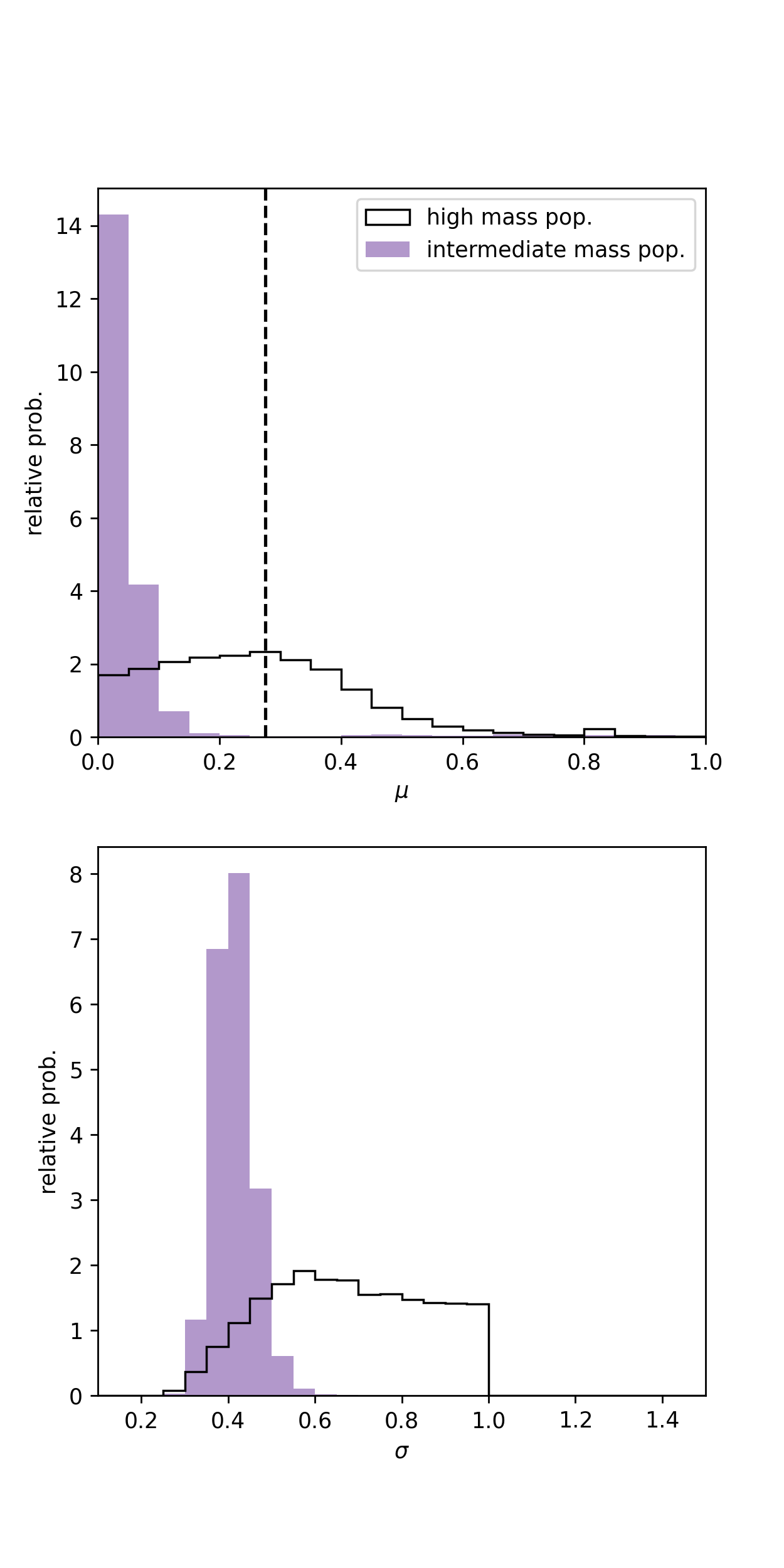}
    \caption{Marginalized posterior distributions of the hierarchical truncated Gaussian fits to the CLS data described in Section \ref{sec:results}, showing the contrast between a preference for a negative (or 0) Gaussian mean eccentricity value among the sub-Jupiter population, and a preferred value of 0.3 for the super-Jupiter population. The dashed line is at the mode value of the high mass population posterior (e=0.275). Takeaway: while we recommend using our hierarchical histogram fit for direct comparison with simulations, the Gaussian fit indicates that the preferred eccentricity ``peak'' location for typical super-Jupiters is at e=0.3.}
    \label{fig:gaussian}
\end{figure}

\begin{figure*}
\includegraphics[trim={2.5cm 0 2.5cm 0}, clip, width=\textwidth]{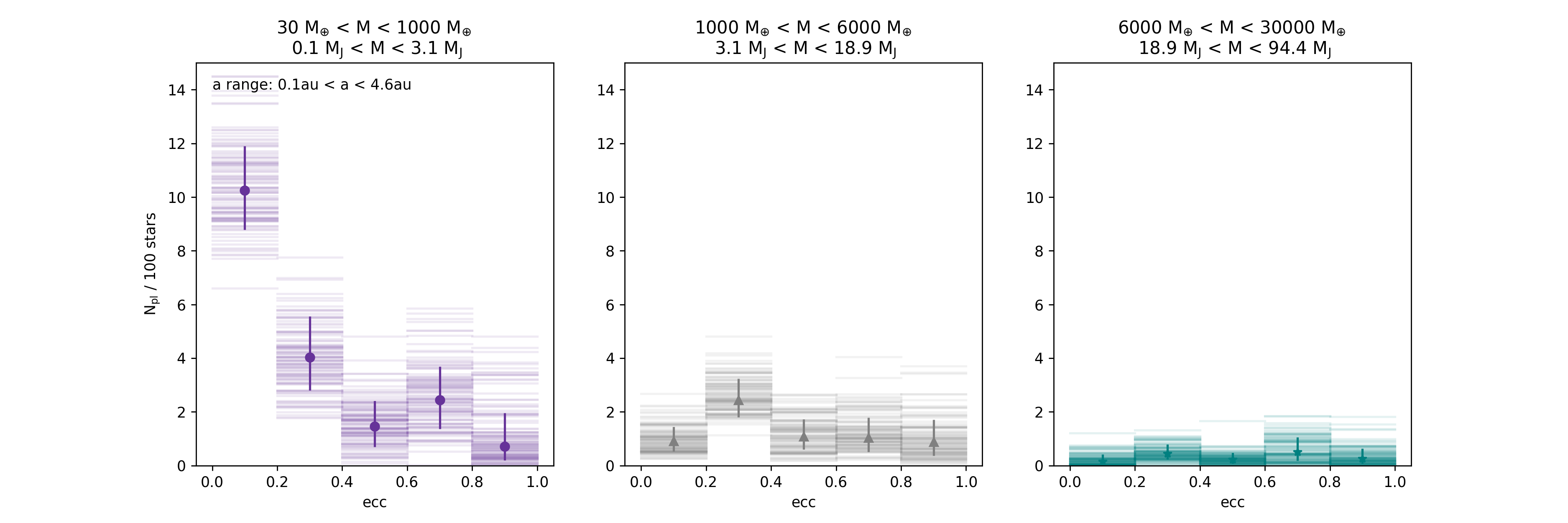}
\caption{Another view of Figure \ref{fig:money}; each panel corresponds to a column of the corresponding panel in Figure \ref{fig:money}. Rather than as a relative probability, the y-axis here is shown as an occurrence rate, highlighting that there are more sub-Jovian than super-Jovian planets per star; occurrence rates in the brown dwarf desert are upper limits. Takeaway: the population-level eccentricity distributions of super- and sub-Jovian planets appear distinct. \label{fig:ecc}}
\end{figure*}

\begin{figure}
\includegraphics[width=\textwidth]{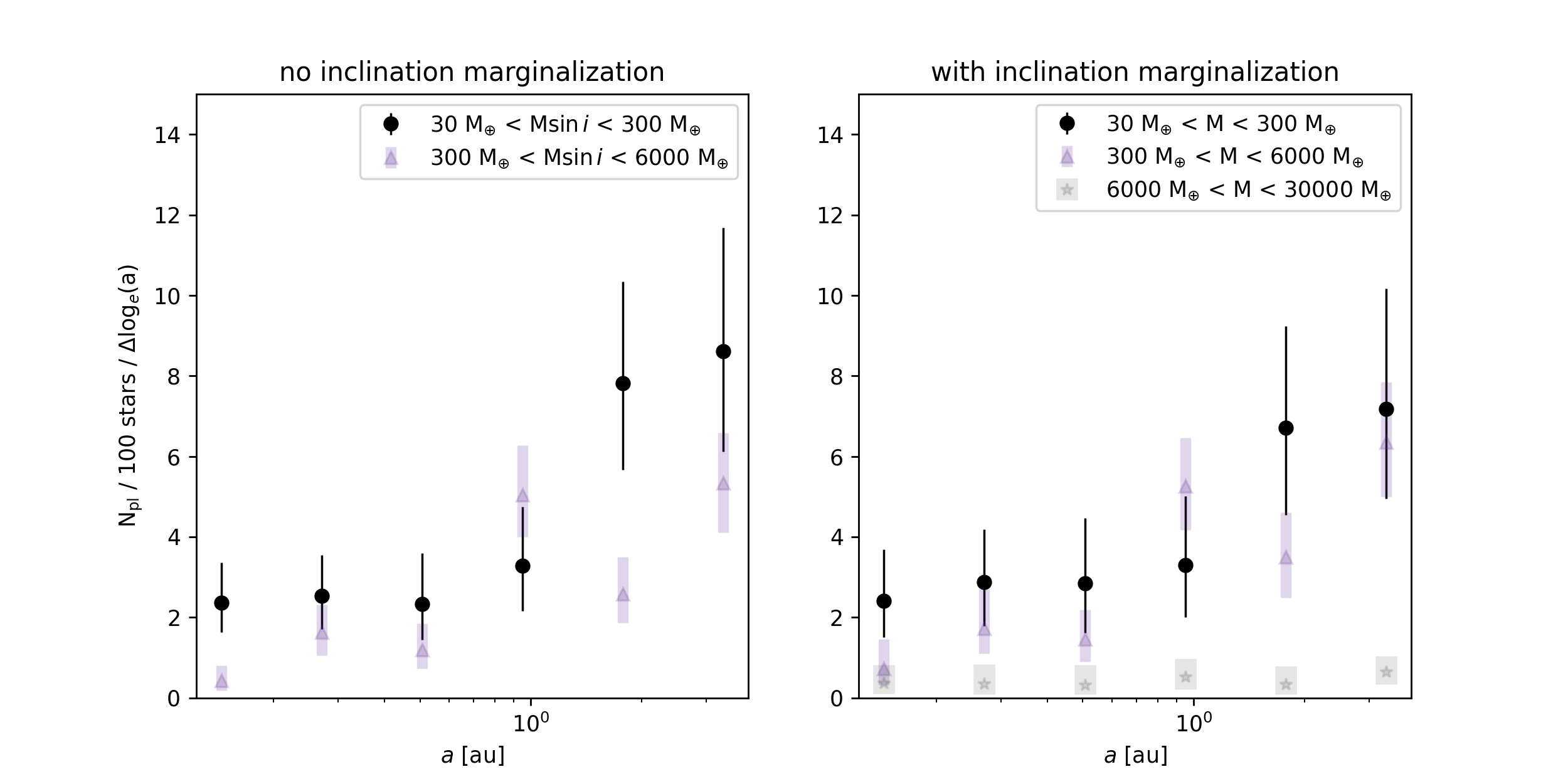}
\caption{Occurrence rate estimates for super-Jovian planets (transparent triangles), sub-Jovian planets (black circles), and brown dwarfs (transparent stars) as a function of semimajor axis. Left: occurrence in log-constant steps of semimajor axis, for two bins in M$\sin{i}$. Right: same as left, but for two bins in \textit{mass}, marginalizing over inclination following the method in the Appendix. This figure can be directly compared to Figure 5 of \cite{Fulton:2021a}, with the caveat stated throughout the text that brown dwarf occurrence rates may be overestimated. Takeaway: we recover the previously identified trend that giant planet occurrence rises with semimajor axis out to $\sim$5~au, both when marginalizing over inclination and not. \label{fig:bj-comp}}
\end{figure}

\begin{figure}
    \centering
    \includegraphics[trim={2cm 0 2cm 0}]{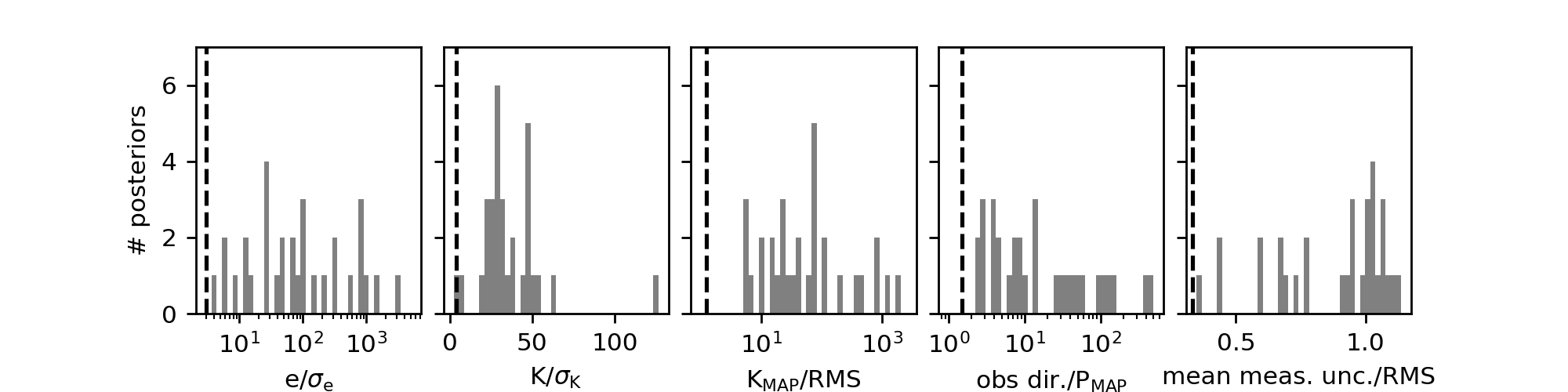}
    \caption{``Plausibility'' metrics for moderately eccentric giant planets, as determined by empirical tests by \cite{Wittenmyer:2019a}. In each panel, we plotted a histogram of a metric of interest (described next) in grey, and the corresponding empirical threshold for plausibility asserted by \cite{Wittenmyer:2019a} in dashed black. They determined that in order for a given eccentric RV planet to be plausible, the following must be satisfied: 1) The fitted eccentricity must be at least 3$\sigma$ from zero (left panel). 2) The fitted velocity amplitude K must be at least four times its own uncertainty: K/$\sigma_{\mathrm{K}} > 4.0$ (second from left panel). 3)  The fitted velocity amplitude must be at least 1.23 times larger than the rms scatter about the fit (middle panel). 4) The fitted period must be less than 1.5 times the total duration of the observations (second from right panel). 5) The rms of the fit must be less than three times the mean measurement uncertainty (right panel). Note that our calculation included fitted jitter values in the calculation of the mean measurement uncertainty. Takeaway: only a single planet fails a single one of these five tests, giving us confidence that the 2-for-1 effect (Section \ref{sec:2for1}) does not explain the trends we report.
    \label{fig:2for1}}
\end{figure}

\section{Discussion \& Conclusion}

\subsection{Summary and Comparison to Previous Work}

In this study, we fit a 3D ($a$, $e$, M) completeness-corrected hierarchical histogram to results from the California Legacy Survey (CLS). We constructed a 3D completeness map using the injection-recovery tests performed by \cite{Rosenthal:2021a}, and used this to fit for the true occurrence rates of giant planets in three mass bins as a function of eccentricity and semimajor axis, focusing on the results we obtained for these planets at the peak of occurrence in semimajor axis (0.1--4.6~au). For the low-mass planets (30--1000 M$_{\oplus}$), at the peak of occurrence, we recovered an eccentricity distribution consistent with monotonically decreasing occurrence as a function of eccentricity. For the high-mass planets (1000--6000 M$_{\oplus}$) at the peak of occurrence, however, we observed a ``peak'' in the eccentricity distribution, which we can identify statistically with 92\% confidence. We also fit a hierarchical truncated Gaussian distribution in place of a hierarchical histogram in order to recover the location of the eccentricity peak with more granularity, finding a preferred value of 0.3.

It is worth emphasizing that, although previous results (e.g. \citealt{gilbert:2025a}, \citealt{Stevenson:2025a}, \citealt{Bowler:2020a}) have recovered a trend of eccentricity increasing with planet mass, in several different semimajor axis and mass regimes, this study is (to our knowledge) the first to show statistical evidence for an eccentricity peak in the cold Jupiter population\footnote{Keep in mind that a monotonically decreasing eccentricity distribution also yields a nonzero average eccentricity.} (however, see \cite{Weldon:2025a}, who also observe a peak but do not attempt to quantify its significance). Our results are similar to those of \cite{Stevenson:2025a}, who recover an increasing average eccentricity as a function of mass for RV-discovered planets, and advocate for splitting the sample into many bins in order to observe trends with sufficient granularity. Our result also bears resemblance to that of \cite{Gupta:2024a} (see their Figure 3), which shows tentative evidence of a ``hole'' in eccentricity-mass space for long-period transiting planets at the lowest eccentricities/highest masses. Although preliminary, this feature appears consistent with our results; we also see a dearth of high-mass planets at low eccentricities and semimajor axes around 1 au (although we have emphasized the pileup of planets at moderate eccentricities instead). Future theoretical work should aim to reproduce all of these observed details.

In the process of computing 3D occurrence rates, we reproduced results from \cite{Fulton:2021a}, which presented occurrence rates as a function of minimum mass and semimajor axis. We translated these occurrence rates to absolute mass, rather than minimum mass, by marginalizing over inclination, although the difference is small. Occurrence rates as a function of mass and semimajor axis can be derived from microlensing studies (see \citealt{Mroz:2023a} for a review), and this aspect of our study is an independent constraint on that same quantity, but eccentricities (and orbital properties in general) are not accessible to microlensing studies.

A key strength of our study is that we perform an empirical completeness correction. The effects of nonuniform completeness have often been ignored in RV eccentricity distribution studies (e.g. \citealt{Kipping:2013a}, \citealt{Weldon:2025a}). We showed (e.g. Figure \ref{fig:completeness}) that RV completeness changes non-negligibly with eccentricity, with fewer than half of planets with eccentricities 0.8-1 being recovered, as opposed to almost 100\% of planets with eccentricities 0-0.2. Because the results we present are estimators of true physical occurrence rates, they can be directly compared to simulation outputs. Planet formation is a process with huge uncertainties, ranging from where planetesimals typically form in a disk to initial planetary multiplicities and subsequent dynamical evolution (see \citealt{Drazkowska:2023a} for a review), but future simulation work (following, e.g. \citealt{Nagpal:2024a} and \citealt{Frelikh:2019a}) could use our results to constrain typical initial multiplicities, masses, and separations after disks dissipate. 

It is possible that the peak we point out in this study has not been quantified before because the parametric models often used to fit eccentricity distributions, specifically the popular Rayleigh and beta distributions, do not have the capability to fit distributions that are nonzero at both 0 and 1 but peaked in the middle. However, there is certainly a trade-off between number of free parameters and recovered information that all modelers must make, and we are not suggesting that RV eccentricity aficionados should never again use the beta or Rayleigh distributions!

\subsection{Implications for Planet Formation}
\label{sec:implications}

Our study is one of a suite of emerging observational clues that constrain the process of giant planet formation. In addition to our work, other salient observational clues include:

\begin{itemize}
    \item eccentricities of warm Jupiters (0.1--1~au) correlate with host star metallicities (e.g. \citealt{Dawson:2013a}, \citealt{Alqasim:2025a})
    \item systems of multiple detected giant planets tend to have lower eccentricities than systems with only one detected giant planet (e.g. \citealt{Rosenthal:2021a}, and tentative evidence from HR 8799 alone for the imaged population from \citealt{Bowler:2020a}).
    \item outer Saturn-mass giant planets tend to correlate with the presence of inner super-Earths, but super-Jupiter systems do not \citep{Lefevre-Forjan:2025a}
    \item if a system has a hot Jupiter, it is more likely to have an outer Jovian companion. In addition, this companion is likely to be more massive than the hot Jupiter itself and to have an elevated eccentricity relative to other Jovian planets \citep{Zink:2023a}
\end{itemize}

There are several ideas in the literature to explain the formation of giant planets. Below, we briefly discuss the mechanism of and evidence for and against each possibility, recognizing that many effects  may simultaneously shape the distribution:

\begin{itemize}
    \item \uline{Formation differences}: although more favored in the literature for planetary mass objects $>10$~au from their host stars (e.g. \citealt{Bowler:2020a}), one idea for explaining differences in eccentricity distribution as a function of mass is that less massive planets formed by core accretion, whereas more massive planets formed by disk instability. This has the advantage of naturally yielding higher eccentricities for planets that formed via collapse, as small angular momentum anisotropies in the disk are expected to be ``frozen in'' as orbital eccentricities after collapse. Because larger objects are expected to be easier to form via collapse, this could explain the elevated eccentricities of more massive objects (see \citealt{Kratter:2016a} for a review). The main problem with this theory is that, although there exist orders of magnitude of uncertainty in disk parameters, the parameters required to form planets down to 3 M$_J$ by disk instability at 1~au are likely unphysical (\citealt{Kratter:2011a}, \citealt{Mordasini:2012a}). It is possible that disk instability plays a role in forming part (or all) of the planetary population we focus on here, but it likely operates much more efficiently for brown dwarf-mass objects at tens of au or farther from their host stars. Real disks may rarely (if ever) collapse at 1~au. For example, \cite{Schib:2025a} performed a population systhesis simulation of planets expected to form via gravitational instability, finding typical companion semimajor axes of $\sim400$au and masses of $\sim30$\mjup{} for primary masses $\sim1M_{\odot}$. They find average eccentricities of $\sim$0.4, but no evidence for a peak at these eccentricities, similar to the distribution of eccentricities of binary stars (e.g. \citealt{Hwang:2022Ma}). Finally, if super-Jupiters typically form by disk instability, while sub-Jupiters form by core accretion, we would not expect the eccentricity-host star metallicity correlation observed by \cite{Alqasim:2025a} to extend to super-Jovian masses. 
    \item \uline{Secular Interactions with Wide Companions}: \cite{Weldon:2025a}\footnote{\cite{Weldon:2025a} also observe an eccentricity peak in their eccentricity distributions using data from the NASA Exoplanet Archive, albeit at (by eye) lower significance than we show here. They do not attempt to quantify the significance of the peak, and also do not perform a completeness correction or hierarchical model to quantify the uncertainty in the distribution.} recently studied the combination of initial formation by core accretion, subsequent planet-planet scattering to moderate eccentricities, and eccentric Kozai-Lidov (EKL) interactions from longer-period stellar companions to reproduce the eccentricity distribution of cold Jupiters (there defined to have semimajor axes 0.8--6~au and masses 0.3--10~M$_J$). This theory, excitingly, appears to reproduce the overall shape we recover for the highest mass planets in our sample here, including the peak at $\sim$0.3. However, the study does not investigate differences in the eccentricity distribution as a function of planetary mass; since we recover very different distributions for super- and sub-Jovian planets, reproducing this difference will be a key test of this theory. EKL does not depend on the mass of the inner companion, so we might not expect population-level eccentricities to change with mass. However, if higher mass planets are more likely to be scattered (perhaps because higher mass planets tend to form in systems with higher multiplicities of higher mass planets), then they may also be more likely to be excited into EKL-induced oscillations. More dedicated simulations, a la \cite{Weldon:2025a}, that aim to reproduce the eccentricity-semimajor axis-planetary mass relationships of cold Jupiters will help clarify whether this is a viable mechanism. Identifying wide binary companions to planets detected by the CLS survey (and other long-baseline RV surveys) will also aid in this effort.
    \item \uline{Tidal effects}: tidal disruption and/or circularization may also play a role in shaping the eccentricity-semimajor axis-mass distribution of typical giant planets. Tidal eccentricity decay is a key ingredient in the high-eccentricity migration theory of hot Jupiter formation, as as discussed in \citealt{Gupta:2024a} (informed by foundational papers \citealt{Petrovich:2015a}, \citealt{Munoz:2016a}, and \citealt{Anderson:2016a}), because lower-mass planets are more likely to cross the Roche radius as they migrate inward, Neptune-mass (sub-Jovian) planets are preferentially removed from the population (forming the ``Neptune desert''), while higher-mass, super-Jovians remain. Although the sample studied in this paper does not include hot Jupiters or hot Neptunes (defined as giant planets with $a<0.1$au), the same effect may be preferentially damping the eccentricities of sub-Jovians. This effect may not fully explain the presence of the eccentricity peak discussed in this paper, but it may partially or fully account for the differences between super- and sub-Jovian eccentricities.
    \item \uline{Secular Planet-disk Interactions}: \cite{Paardekooper:2023a} provide an excellent recent overview of recent progress in planet-disk interaction studies (see their Section 3.9). For planets capable of opening a gap in the disk ($>$ 1 \mjup{}), \cite{Bitsch:2013a} found through simulations that planet-disk interactions typically reduce eccentricities, but \textit{can} increase eccentricities when either a) the planet is massive enough to open a deep gap, or b) the planet is sufficiently inclined that the disk excites the planet into Kozai-Lidov oscillations. Subsequent studies performing detailed 2- and 3D simulations (e.g. \citealt{Duffell:2015a}, \citealt{Lega:2021a}) found that eccentricity excitation due to interactions between massive gap-opening planets and their disks was possible, but only saw excitation up to small maximum values in simulations (0.07 and 0.25, respectively). Another idea, proposed by \citet{Teyssandier:2016a}, is that planets that migrate into the central disk cavity may be excited to eccentricities as large as 0.4 \citep{Debras:2021a} by Lindblad resonances excited in the disk. More simulations in this area, particularly investigating the impact of planet semimajor axis and mass, are needed to conclusively determine whether this mechanism is a viable explanation for the trends we observe.
    \item \uline{Planet-planet Scattering and Giant Impacts}: as nicely summarized by \cite{Weldon:2025a}, core accretion plus subsequent planet-planet scattering alone is unlikely to be responsible for the cold Jupiter population. While the final eccentricity distributions produced by scattering are very sensitive to initial conditions, they tend to follow a Rayleigh-distribution (e.g. \citealt{Chatterjee:2008a}, \citealt{Juric:2008a}, \citealt{Zhou:2007a}), which is inconsistent with the distribution shapes we find here, which are nonzero at both 0 and peak at 0.3 (although, intriguingly, the resulting distributions do often peak at $\sim$0.4; see e.g. \citealt{Juric:2008a}). However, as \cite{Frelikh:2019a} pointed out, at separations of $\sim$1~au, the impact cross sections of giant planets equal the scattering cross sections at lower eccentricities for lower masses (at constant planetary radius). Therefore, because giant planet radius does not change substantially with mass, higher eccentricities can be achieved before collisions occur. Thus, if it is common to form systems of multiple giant planets, these planets may scatter each other up to moderate eccentricities before they collide. This theory naturally explains the higher eccentricities of more massive planets at separations of $\sim$few au; although in an individual scattering event, a less massive object will be scattered to a higher eccentricity than a more massive object, at a population level, eccentricity can be interpreted as a tracer of a merger, so planets with higher eccentricities are also expected to 1) be more massive, 2) have formed in systems with more giant planets (hence the correlation with eccentricity and metallically, a la \citealt{Alqasim:2025a}), and 3) disrupt the orbits of (or prevent the formation of altogether) inner small planets (explaining the trend of \citealt{Lefevre-Forjan:2025a}). 
\end{itemize}

It is interesting to consider the results of \cite{Zink:2023a} in the context of these theories. In general, planet-planet scattering as a pathway to hot Jupiter formation appears to explain the correlation between hot Jupiters and outer Jovian companions, which may have formed closer together and then scattered off each other. The elevated eccentricities of these outer companions could either be a direct result of the scattering event itself, slightly damped from the value post-scattering due to collisions, or oscillating around the value post-scattering due to secular interactions with even wider separated binary companions. 

\subsection{Future Work \& Final Remarks}

Future work on the eccentricity distributions of the CLS sample could develop a more robust completeness model by performing more granular injection-recovery tests and/or fitting a parametric model to avoid the assumption that completeness is uniform across histogram bins. Another interesting follow-up avenue would be further work investigating where the eccentricity peak appears in parameter space. In particular, we did not explore the location or shape of the eccentricity peak as a function of mass. Combining results from multiple long-baseline RV surveys might give the necessary sample size to observe these more subtle trends. 

The near-future Gaia Data Release 4 (DR4) will likely be sensitive to thousands of ``typical'' Jovian planets around bright stars (\citealt {Perryman:2014a}, \citealt{Lammers:2026a}),  which should reveal much more granular eccentricity trends that what we have been able to find here. However, the lessons learned in this paper will be important to keep in mind for the Gaia DR4 release; in particular, the methodology developed in the appendix may come in handy for resampling  eccentricity posteriors derived in a different fitting basis. 

Although Jupiter, at 5.2~au, is slightly outside of the semimajor axis limits of our sample, it is interesting to consider its properties in the context of these results. Jupiter's small eccentricity (0.05) would fall in the eccentricity bin with the highest median posterior occurrence rate for its mass. In other words, in terms of eccentricity, semimajor axis, and mass, Jupiter does appear ``typical'' among its peer exoplanets. However, for a more massive Jupiter ($>$3~M$_{\rm J}$) at the peak of occurrence, such a small eccentricity might be more unusual. If planets more massive than Jupiter really are typically eccentric, the fact that Jupiter is massive, but not too massive, may be an intriguing precondition for the stability (and perhaps habitability) of Earth. 

\begin{acknowledgments}

S.B. would like to thank Eric Nielsen, Bill Roberson, Megan Bedell, Brendan Bowler, Marvin Morgan, Greg Gilbert, Steven Giacalone, Kendall Sullivan, Nick Choski, Malena Rice, and Devin Cody for helpful discussion (and burritos and emotional support). S.B. would also like to thank Beyonc\'e Knowles-Carter for the phrase ``always stay gracious best revenge is your paper.''

S.B. wishes to acknowledge her status as a settler on the ancestral lands of the Muwekma Ohlone tribe, and to recognize that the literature astronomical observations used in this paper were only possible because of the dispossession of Maunakea from Kan\={a}ka Maoli. 

All of the authors wish to thank the anonymous referee for thoughtful comments that improved this paper.

J.J.W.\ is supported by NASA Grant 80NSSC23K0280.

\end{acknowledgments}

\begin{contribution}
S.B. and J.W. conceived of the study, and S.B. performed the analysis, made the plots, and wrote the manuscript. J.W., R.M.-C., B.M., and R.R. provided advice, assistance with interpretation, and comments on the draft. B.J.F. performed the foundational injection-recovery tests that much of this work is based on and assisted with questions about methodology. 

\end{contribution}

\software{\texttt{numpy} \citep{harris2020array}, \texttt{pandas} \citep{pandas}, \texttt{matplotlib} \citep{Hunter:2007aa}, \texttt{corner} \citep{corner}, \texttt{scipy} \citep{scipy}, \texttt{astropy} (\citealt{astropy:2013}, \citealt{astropy:2018}, \citealt{astropy:2022}), \texttt{radvel} \citep{Fulton:2018a}}

\appendix

\section{Mathematical Formalism}
\label{sec:appendix1}

The statement of the problem is as follows: we have N planets that were discovered (or recovered) in a uniform survey, each with its own (marginalized) joint posterior distribution over eccentricity $e$, semimajor axis $a$, and minimum mass \msini{}, and we want to construct a posterior distribution over the parameters of a model representing the underlying true occurrence rate of planets, $\Gamma_{\theta}(\boldsymbol{\omega})$, where $\boldsymbol{\omega}$ represents the vector: $\boldsymbol{\omega}$ $:=$ ($e$, $M$, $a$), and  $\Gamma_{\theta}$ is a function we choose that depends on parameters $\boldsymbol{\theta}$. We construct a posterior distribution over $\boldsymbol{\theta}$ given our survey results, which we can use Bayes theorem to write as:

\begin{equation}
\begin{split}
p( \boldsymbol{\theta} | \{\boldsymbol{V}\} ) \propto p(\{\boldsymbol{V}\} |  \boldsymbol{\theta} )p( \boldsymbol{\theta} )
\end{split}
\end{equation} where $\{\boldsymbol{V}\}$ is the set of all observations taken by the survey, and $p(\boldsymbol{\theta})$ defines the prior on the parameters of the occurrence model $\Gamma_{\theta}$.

Following \cite{Foreman-Mackey:2014a}, we define $\Gamma_{\theta}(\boldsymbol{\omega})$ as:

\begin{equation} \label{eq:gamma}
    \Gamma_{\theta}(\boldsymbol{\omega}) = \frac{d^3N}{de\,dM\,da}
\end{equation} such that integrating 

\begin{equation}
    \int_{\boldsymbol{\omega}_i} \Gamma_{\theta}(\boldsymbol{\omega}) d\boldsymbol{\omega}  = N_i
\end{equation} over the boundaries of the $i$th bin gives the true number of planets in that bin (according to our model), N$_i$.

% Before continuing, we can introduce a statistical ``hack;'' although the occurrence rates for the CLS sample given in \cite{Fulton:2021a} are reported in terms of M$\sin{i}$, this is not ideal. We would like to be able to provide occurrence rates in terms of absolute mass if possible. To statistically sidestep this difficulty, we can notice that because radial velocities alone do not constrain inclination $i$, the posterior over this parameter will be exactly our prior. We can assume that the orbital angular momentum vector has a uniform distribution over a sphere, which is equivalent to assuming that the $\cos{i}$ ~ $\mathcal{U}$(-1,1). We can then obtain random samples from $\cos{i}$ posteriors by taking random samples from $\mathcal{U}$(-1,1), then converting each of our \msini{} samples to samples in $M$ by dividing by the sampled values of $\sin{i}$. 

% Effectively, this ``hack'' allows planets with a given value of M$\sin{i}$ to ``float'' to higher mass bins in accordance with its expected distribution of true masses. One caveat here is that we are approximating that completeness to values of \msini{} below 30 M$_{\oplus}$ is 0 in our semimajor axis range of interest (beyond 0.1~au); this likely causes occurrence rates in the smallest \msini{} bins to be slightly underestimated. Future work should develop a more fine-grained completeness model lifting this approximation. 

% Having defined our hack, we can now continue by defining our model, 

An immediate complication is that individual orbital parameter posteriors are given in terms of M$\sin{i}$, not M. To statistically sidestep this difficulty, we can introduce prior information about the distribution of inclinations; at a given value of $\boldsymbol\omega$, we can assume that the orbital angular momentum vectors of all planets are drawn from a uniform distribution over a sphere, which is equivalent to assuming that $\cos{i}\sim\mathcal{U}$(-1,1). Therefore, although we define occurrence in $e$, $a$, $M$, and $i$, we can compute the occurrence of a planet with measured properties ($a$, $e$, $M_{\rm min}$) as:
\begin{equation}
\begin{split}
\label{eq:marginal}
\Gamma_{\theta}(a, e, M_{\rm min}) &= \int_{i=0}^{i=90}\Gamma_{\theta}(a, e, M(i), i)di\\
&= \int_{i=0}^{i=90}\Gamma_{\theta}(a, e, \frac{M_{\rm min}}{\sin{i}}, i)di\\
&= \int_{i=0}^{i=180}\Gamma_{\theta}(a, e, \frac{M_{\rm min}}{\sin{i}})\frac{\sin{i}}{2}di\\
\end{split}
\end{equation}

Equipped with our statistical hack, we now define a corresponding \textit{completeness}, Q, in $e$, $a$, and M$_{\mathrm{min}}$ (which we assume is known a priori), and write down a full estimator for the observed occurrence rate for a planet with measured properties a$_j$, e$_j$, M$\sin{i}_j$:

\begin{equation}
\hat{\Gamma}_{\theta}(a, e, M_{\rm min}) = Q(a, e, M_{\rm min}) \Gamma_{\theta}(a, e, M_{\rm min}), 
\end{equation} where $\Gamma_{\theta}(a, e, M)$ is the true underlying occurrence rate of planets-- what want to infer. 

Having defined our model and estimator, we can now follow \cite{Youdin:2011a} to derive a likelihood that assumes no observational errors, $p(\{{\boldsymbol{\omega}}\} | \theta)$, or equivalently, a likelihood that assumes we have perfect knowledge of the orbital parameters of each planet in our sample. \cite{Youdin:2011a} noticed that the likelihood of observing a certain number of planets $\hat{\rm N}_{i}$ in a particular bin of parameter space (which we will call bin $i$) is Poisson distributed, and therefore that the likelihood of observing the results of a survey is the product of the Poisson likelihoods for each bin. This gives a likelihood of (with constant factors removed):

\begin{equation}
\begin{split}
    p(\boldsymbol{\omega} | \boldsymbol{\theta}) &\propto exp \left[{-\int \hat{\Gamma}_{\theta} (\boldsymbol{\omega}) d\boldsymbol{\omega}} \right] \prod_{i=0}^{B} (\hat{\Gamma}_{\theta} (\boldsymbol{\omega}_i))^{B_{i}}\\
    & = exp \left[{- \int \hat{\Gamma}_{\theta} (\boldsymbol{\omega}) d\boldsymbol{\omega}} \right] \prod_{i=0}^{N} (\hat{\Gamma}_{\theta} (\boldsymbol{\omega}_i)),
\end{split}
\end{equation} where in the first line we are summing over bins, and in the second line we are equivalently summing over observed planets. As pointed out by \cite{Foreman-Mackey:2014a}, this procedure is analogous to making a histogram; the maximum likelihood solution here, if the completeness values were all 1, would be exactly a histogram.

If the planets discovered by our sample had perfectly known values of $e$, $M$, and $a$, we would be done. However, this is not the case for our sample of interest, where posterior estimates of planet orbital parameters often have noneligible error bars. To account for this complication, we instead define our likelihood not as the joint probability of observing a set of perfectly defined \textit{orbital parameters} given the occurrence model, but as the joint probability of observing each set of \textit{RV data points} given the occurrence model. Taking this complication into account, our new hierarchical likelihood is:

\begin{equation}
\label{eq:hier}
\begin{split}
p(\{\boldsymbol{V}\} | \boldsymbol{\theta}) = \prod_{i=0}^{N}\int p(\boldsymbol{V_i} | \boldsymbol{\omega_i}) p(\boldsymbol{\omega}_i | \boldsymbol{\theta}) d\boldsymbol{\omega_i}
\end{split}
\end{equation} where $\boldsymbol{V_i}$ is the RV timeseries measurements for the $l$th star in the sample, and $\{\boldsymbol{V}\}$ is the set of all timeseries measurements in the sample.

We next follow \cite{Hogg:2010a} and \cite{Foreman-Mackey:2014a} to rewrite the above in terms of individual planet posterior distributions\footnote{These are assumed to be independent of one another.}, $p(\boldsymbol{\omega_i}|\boldsymbol{V}_i,\boldsymbol{\alpha})$\footnote{Note that throughout this derivation, we remove constant factors that do not depend on our free parameters $\boldsymbol{\theta}$ for clarity}. $\boldsymbol{\alpha}$ is introduced here to indicate that each planet posterior was computed assuming an \textit{intermediate} prior, itself a function of parameters $\boldsymbol{\alpha}$. This substitution will allow us to reuse samples from precomputed posteriors for each individual star, rather than computing the entire joint likelihood for all stars and all data points at each MCMC step. In this ``intermediate'' step, $p(\boldsymbol{V_i} | \boldsymbol{\omega_i})$ can be recast as a likelihood:

\begin{equation}
\label{eq:bayes-int}
\begin{split}
    p(\boldsymbol{\omega_i} | \boldsymbol{V_i},\boldsymbol{\alpha} ) \propto p(\boldsymbol{V_i} | \boldsymbol{\omega_i}) p(\boldsymbol{\omega_i}|\boldsymbol{\alpha})
\end{split}
\end{equation} where, recall, $\boldsymbol{\alpha}$ is the vector of parameters that describe the intermediate prior. Because $p(\boldsymbol{V_i} | \boldsymbol{\omega_i})$ is independent of the prior, it does not depend on $\boldsymbol{\alpha}$. Now, we can substitute Equation \ref{eq:bayes-int} into Equation \ref{eq:hier} to obtain: 

\begin{equation}
\begin{split}
p(\{\boldsymbol{V}\} | \boldsymbol{\theta}) = \prod_{i=0}^{N} \int \frac{p(\boldsymbol{\omega}_i | \boldsymbol{\theta})}{p(\boldsymbol{\omega}_i |\boldsymbol{\alpha})} p(\boldsymbol{\omega_i}|\boldsymbol{V_i},\boldsymbol{\alpha}) d\boldsymbol{\omega_i}.
\end{split}
\end{equation} The final step is to use the Monte Carlo integral approximation to substitute a list of samples drawn from the distribution (e.g. an MCMC chain) in for $p(\boldsymbol{\omega_i}|\boldsymbol{V_i},\boldsymbol{\alpha})$, the posterior of parameters for the $i$th planet, following \cite{Hogg:2010a} Equation 8. 

Making this final substitution, our approximate hierarchical likelihood is:

\begin{equation}
\begin{split} \label{eq:money}
p(\{\boldsymbol{V}\} | \boldsymbol{\theta}) &\propto \prod_{i=0}^{N}  \frac{1}{N_i} \sum_{n=1}^{S_i} \frac{p(\boldsymbol{\omega}_i^{(n)} | \boldsymbol{\theta})}{p(\boldsymbol{\omega}_i^{(n)} |\boldsymbol{\alpha})} \\
&= exp \left[{- \int \hat{\Gamma}_{\theta} (\boldsymbol{\omega}) d\boldsymbol{\omega}} \right] \prod_{i=1}^{N} \frac{1}{N_i} \sum_{n=1}^{S_i} \frac{\hat{\Gamma}_{\theta}(\boldsymbol{\omega}_i^{(n)})}{p(\boldsymbol{\omega}_i^{(n)}|\boldsymbol{\alpha})}
\end{split}
\end{equation} where $S_i$ is the number of samples drawn from the $i$th posterior, and $\omega_k^{(n)}$ are precomputed posterior samples. 

When this method has been used previously, generally it has been assumed that the intermediate prior ${p(\boldsymbol{\omega}_k^{(n)}|\alpha)}$ is uniform, or otherwise unimportant. In this study, for reasons that will be clarified in Section \ref{sec:appendix2}, the intermediate prior is a) nonuniform, and b) improper, and therefore impossible to compute analytically. In Section \ref{sec:appendix2}, we develop an approximate method to mitigate this difficulty. This turns out to have a negligible effect for the present study, but may be useful for future work where the intermediate priors have more impact on the individual planet posteriors.

\subsection{Functional Forms}

So far, we have not defined a functional form for our occurrence rate model $\Gamma_{\theta}$. The functional form we use in this paper is a piecewise constant step function, what we call a hierarchical histogram; following \cite{Foreman-Mackey:2014a}, this is defined as:

\begin{equation}
\begin{split}
    \Gamma_{\boldsymbol{\theta}}(\boldsymbol{w})= \begin{cases} \theta_1 & \boldsymbol{w} \in \Delta_1 \\ \theta_2  & \boldsymbol{w} \in \Delta_2 \\ \cdots & \\ \theta_B  & \boldsymbol{w} \in \Delta_B \\ 0 & \text { otherwise }\end{cases}
\end{split}
\end{equation} where the bins $\Delta_B$ are defined in the 3D space (a, e, M); a priori, we divide up the 3D parameter space into $B_e$ eccentricity bins, $B_a$ semimajor axis bins, and $B_{M}$ minimum mass bins, for a total of $B_eB_aB_M := B$ bins. These bins set the boundaries of a 3D histogram, with the histogram bin ``heights,'' $\theta_B$, being the free parameters in our model which describe the true occurrence rate of planets in that $e$, $a$, and $M$ bin. Plugging this parametrization into Equation \ref{eq:marginal} yields:

\begin{equation}
\begin{split}
\label{eq:msini}
\Gamma_{\theta}(a, e, M_{\rm min}) &= \sum_{j}\int_{\sin^{-1}\frac{M_{\rm min}}{M_j}}^{\sin^{-1}\frac{M_{\rm min}}{M_{j+1}}}\Gamma_{\theta}(a, e, M_j)(\sin{i})di\\
&= \sum_{j}\Gamma_{\theta}(a, e, M_j)\left( \sqrt{1 -  \frac{M_{\rm min}^2}{M_{j+1}^2}} - \sqrt{1 - \frac{M_{\rm min}^2}{M_{j}^2}}\right)\\
\end{split}
\end{equation}

where the sum over $j$ indicates a sum over the boundaries of the $j$th mass bin; M$_j$ is the lower boundary of the $j$th bin, M$_{j+1}$ is the upper boundary, and $\Gamma_{\theta}(a, e, M_j)$ is the occurrence rate in that bin.

Although equation \ref{eq:msini} looks complicated, it boils down to a simple idea: transforming an occurrence rate in mass to an occurrence rate in \msini{} involves multiplying each mass occurrence rate by probability it is actually in that mass bin, given its \msini{}.

\section{An Approximate Method to ``Undo'' the Effective Priors Set by Sampling in a Non-physical Basis}
\label{sec:appendix2}

In this study, we used precomputed MCMC chains which estimate the marginalized posterior over $e$, $a$, and \msini{} for each detected planet in the CLS survey \citep{Rosenthal:2021a}. 
However, the posteriors were not computed in the ($e$, $a$, and \msini{}) basis directly, but in the basis ($\sqrt{e}cos{\omega}$, $\sqrt{e}sin{\omega}$, $P$, $K$, $T_C$), \footnote{This is different from the prior basis that was stated in \cite{Rosenthal:2021a} and \cite{Fulton:2021a}, which asserts that fits were performed in a basis using $\log{P}$, not $P$. However, the public versions of the set up files used to run the fits, available at \href{https://github.com/leerosenthalj/CLSI/tree/master/radvel_setup_files}{this link}, all use a basis including $P$. We assume that this was the basis used in the chains shared by Fulton et al (private communication). In practice, the entire prior resampling process does not affect our results at all, so the difference does not matter.} with improper uniform priors assumed on all parameters. This does \textit{not} translate to priors that are uniform in $a$ and \msini{} \footnote{Although it does for $e$, when $e<$1; see \cite{Lucy:1971a}.}, and therefore the denominator, $p(\boldsymbol{\omega}_i^{(n)}|\boldsymbol{\alpha})$, in Equation \ref{eq:money}, cannot be ignored.

In this section, we develop an approximate method for determining and ``undoing'' the effective priors that were applied to our available posterior samples. Posterior samples in $e$, $K$, and $P$ are converted to samples in $e$, \msini{}, and $a$ via the pair of equations (from Keplerian orbital motion):

\begin{equation}
\begin{split}
    \frac{M\sin{i}}{M_{\rm Jup}} &= \left( \frac{M_*}{M_{\odot}} \right)^{\frac{2}{3}} \frac{K}{28.4325 \, \rm m \, s^{-1}} \left(\frac{P}{\rm yr}\right)^{\frac{1}{3}} \sqrt{1 - e^2}\\
    \frac{a}{\rm au} &= \left(\left(\frac{P}{\rm yr}\right)^2 \left(\frac{M_*}{M_{\odot}}\right)\right)^{1/3}.
\end{split}
\end{equation} We can determine the effective (joint) prior on \msini{}, $e$, and $P$ by applying results from the algebra of random variables. Specifically, we assume that $K$ and $P$ are uniform random variables drawn from improper priors defined on [0, $\infty$), $e$ is a uniform random variable defined on [0,1), and $M_*$ is a constant.\footnote{Strictly speaking we should assume a Gaussian prior on $M_*$, using the mean and standard deviation of the stellar masses derived from spectroscopy in \cite{Rosenthal:2021a}, but practically speaking these priors are so narrow that the difference is negligible.} In general, if X is a  random variable drawn from the probability distribution p$_{\rm X}$(x), and Y is some function of X, Y:=g(X), then Y will follow the distribution:

\begin{equation}
\begin{split}
    f_Y(y) =  f_X(g^{-1}(y))\left|\frac{d}{dy} g^{-1}(y)\right|
\end{split}
\end{equation} \citep{Springer:1979a}. Notice that if f$_X$(x) is uniform, the first multiplied term reduces to a constant, and we recover the process of inverse transform sampling, commonly used in rejection sampling or nested sampling processes to transform uniform random samples to random samples from another distribution. One problem we run into applying this transformation is that our priors on $K$ and $P$ are improper (i.e. they are defined over an infinite range and do not sum to 1). To overcome this difficulty, we defined values K$_{max}$ and P$_{max}$ to be twice the maximum sampled value of the K and P posteriors, respectively, reasoning that the sampled posteriors would have been unchanged under uniform priors with these maximum values. With this simplification, the prior transformation can now be computed analytically. The effective prior on $a$, f$_a$(x), is given by:

\begin{equation}
\begin{split}
    f_a(x) \propto x^{1/2}\\
\end{split}
\end{equation} and the effective prior on \msini{}, f$_M$(x), is given by:

\begin{equation}
\begin{split}
    A &:= P_{max}^{-2/3}\\
    f_M(x) &\propto \frac{\rm sin^{-1}\left({\sqrt{A}x}\right)}{2 \sqrt{A}}  +  \frac{x}{2}\sqrt{1 - Ax^2}
\end{split}
\end{equation}

\begin{figure*}
    \includegraphics[]{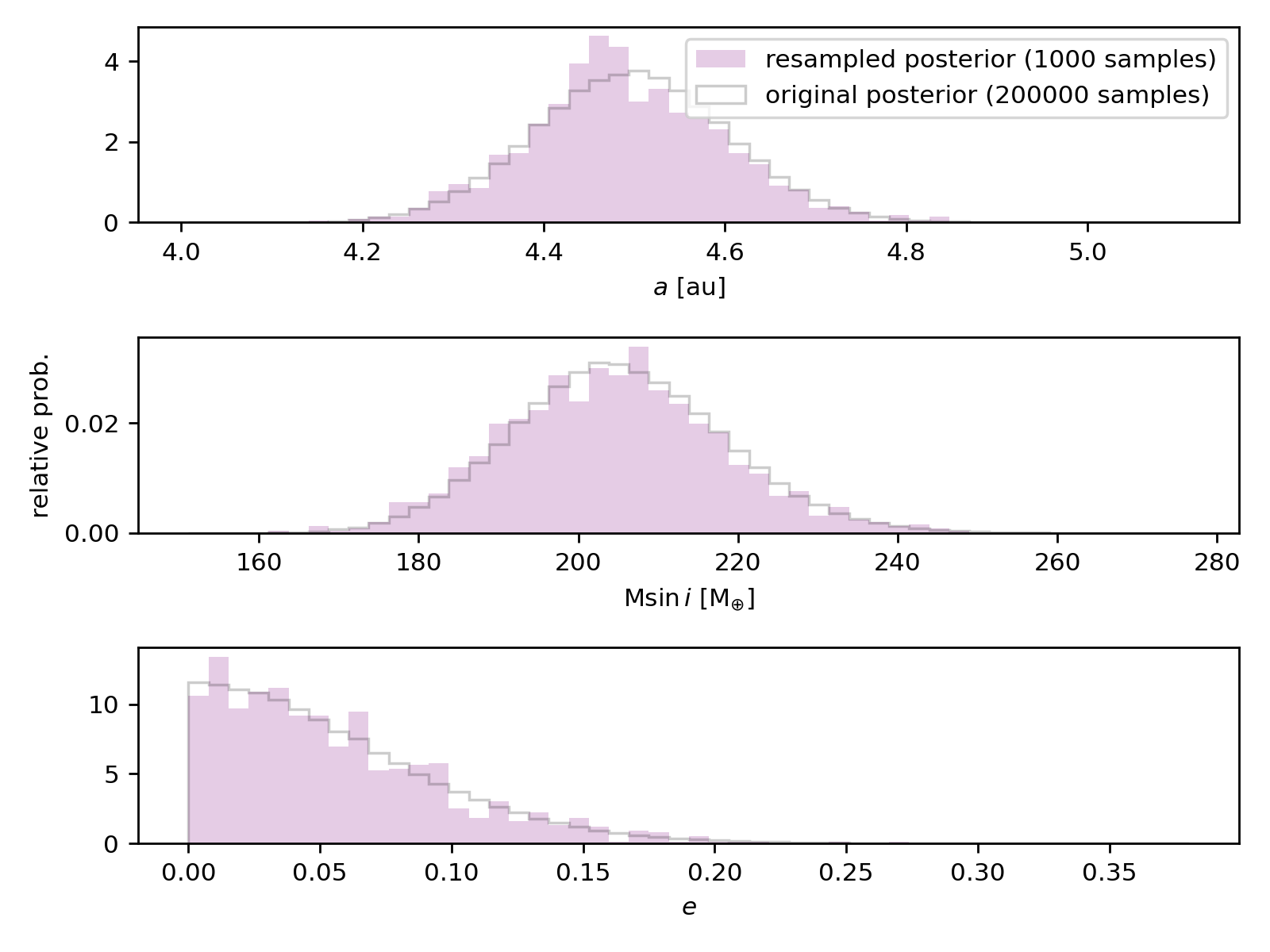}
    \caption{Original marginalized posterior, obtained via uniform priors on $P$ and $K$ (gray unfilled histogram), and resampled posterior (purple shaded histogram), obtained by importance resampling. The resampled posterior estimates what the posterior \textit{would} have looked like assuming uniform sampling in $\log{\rm M\sin{i}}$ and $\log{a}$. The posterior is virtually unchanged by the act of posterior resampling because it is quite prior-independent. \label{fig:resampling-good}}
\end{figure*}

\begin{figure*}
    \includegraphics[]{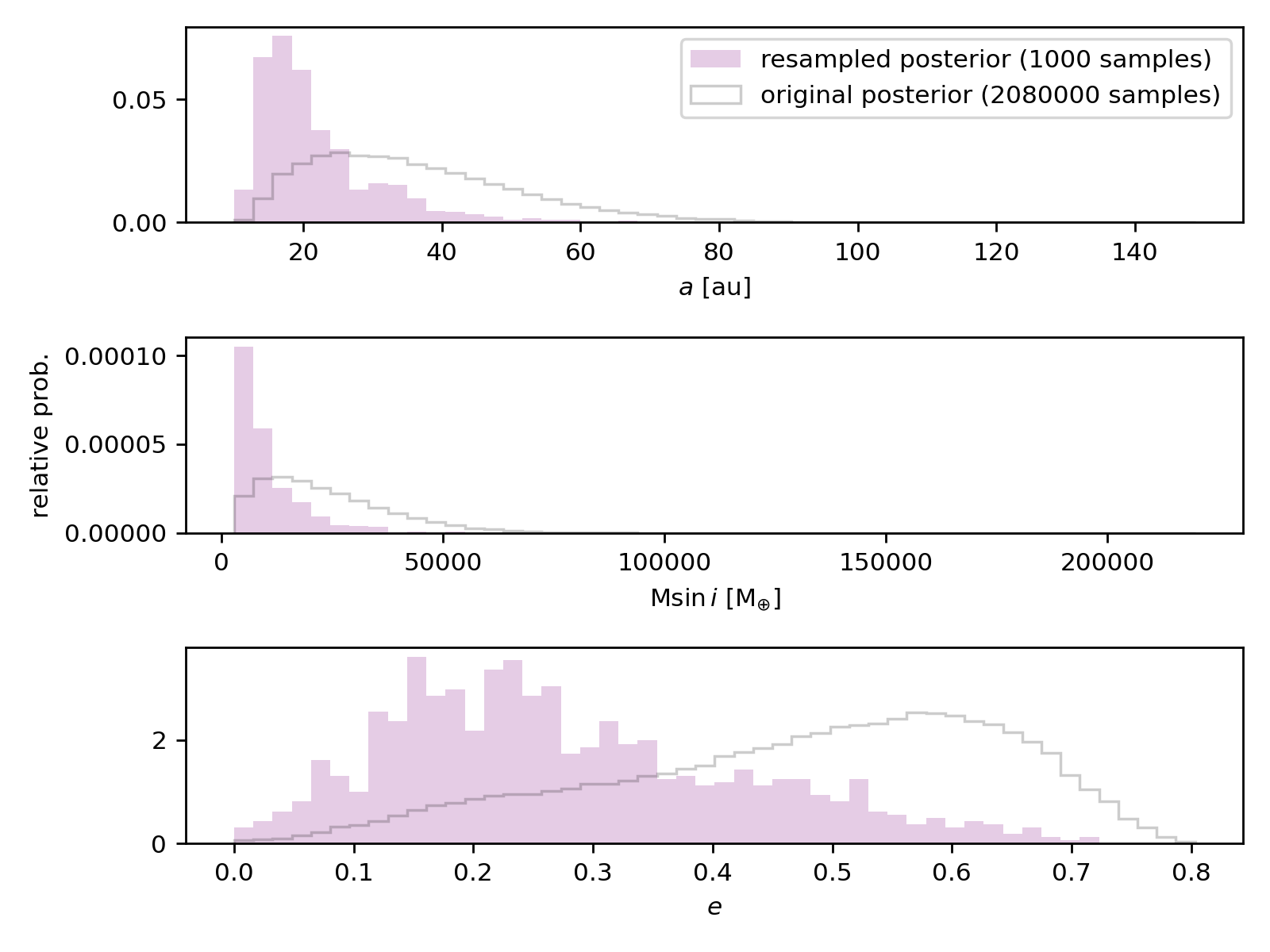}
    \caption{Same as figure \ref{fig:resampling-good}, but for an object in the CLS sample with an incomplete orbit. This posterior depends sensitively on the prior assumed, and therefore the act of prior resampling significantly changes the derived eccentricity posterior. Because the posterior has exhibits covariances between $e$, \msini{}, and $a$, the act of prior resampling affects the marginalized eccentricity posterior, even though the 1D prior on eccentricity has is unchanged. \label{fig:resampling-bad}}
\end{figure*}

At this point, we have determined the ``effective'' priors on \msini{} and $a$, and we can now use \textit{importance sampling} \citep[e.g.]{Jiang:2022a} to resample the posteriors\footnote{Note that we could equivalently substitute $p_M(M\sin{i})p_a(a)$ in for $p(\boldsymbol{\omega}_k^{(n)}|\alpha)$ directly in Equation \ref{eq:money} when computing our hierarchical likelihood. However, we prefer to use importance sampling so that we can also observe the impact of the effective prior on each individual object in the sample.}. This procedure mirrors the one used in \cite{Hogg:2010a} to derive the approximation to the hierarchical likelihood. In terms of the variables used in Section \ref{sec:appendix1}, we have samples $\boldsymbol{\omega}_k^{(n)}$, which are drawn from the posterior under the intermediate prior $\alpha$. This importance resampling method reduces to a weighted random draw, with the  weights given by:

\begin{equation}
\begin{split}
    w_i \propto \frac{p_{\rm new \, prior}(M\sin{i},a)}{p_M(M\sin{i})p_a(a)} \frac{1}{\sum{w_i}}.
\end{split}
\end{equation} This procedure is approximately the same as re-running the entire MCMC chain using a different prior. Figures \ref{fig:resampling-good} and \ref{fig:resampling-bad} show this method in action. 
Figure~\ref{fig:resampling-good} shows a complete orbit typical of our sample. In this case, the data are highly constraining, so the result is prior-independent and does not depend on resampling. Figure~\ref{fig:resampling-bad} demonstrates that in the case of an incomplete orbit, resampling is critical for removing the bias introduced by the effective prior. However, this particular case does not significantly impact our occurrence results, since the majority of the posterior weight in semimajor axis places this system well outside the bounds of our sample.

Some important limitations of this method include:

\begin{itemize}
    \item if the new prior extends to regions of parameter space where the old prior does not (i.e. the prior was 0), the posterior resampling method may not produce valid results.
    \item because we are drawing a random subset of samples from the existing posterior, this method is only valid when the number of samples needing to be drawn is much less than the existing number of samples. This is fortunately true in our case; typical chain lengths are $\sim$ millions, and we only need $\sim$1000 samples per posterior in order to produce accurate hierarchical posterior samples.
\end{itemize}

As stated above, performing this prior resampling has a negligible effect on the posteriors in our parameter space of interest (0.1-5 au, 30-6000 M$_{\oplus}$ minimum mass). However, as Figure \ref{fig:resampling-bad} illustrates, it has a \textit{significant} effect on individual posteriors for incomplete orbits, and we encourage future analyses using the approximate hierarchical Bayesian methodology of \cite{Hogg:2010a} to account for the important effects of non-uniform intermediate priors.

\bibliography{biblio}{}
\bibliographystyle{aasjournalv7}

\end{document}